\documentclass[aps,pra,superscriptaddress]{revtex4}

\usepackage{latexsym}
\usepackage{amsmath}
\usepackage{graphicx}

\begin{document}
%


\title{Topology Induced Macroscopic Quantum Coherence in
Josephson Junction Networks}

\author{G. Giusiano}
\author{F. P. Mancini}
\author{P. Sodano}
\affiliation{Dipartimento di Fisica and Sezione I.N.F.N., Universit\`a
di Perugia, \\ Via A. Pascoli
Perugia, I-06123,  Italy}
\author{A. Trombettoni}
\affiliation{I.N.F.M. and Dipartimento di Fisica, Universit\`a
di Parma,\\ parco Area delle Scienze 7A
Parma, I-43100, Italy}

\date{\today}


\hyphenation{periodically pe-ri-od-i-cal-ly}

\begin{abstract}
We argue that Josephson junction networks may be engineered to allow for
the emergence of new and robust quantum coherent states. 
We provide a rather intuitive argument showing how
the change in topology may affect the quantum properties 
of a bosonic particle hopping on a network.
As a paradigmatic example, we analyze in detail the quantum and
thermodynamic properties of non-interacting bosons hopping on
a comb graph. We show how to explicitly compute 
the inhomogeneities in the distribution of bosons along the 
comb's fingers, evidencing the effects of the topology induced spatial 
Bose-Einstein condensation characteristic of the system.
We propose an experiment enabling to detect 
the spatial Bose-Einstein condensation 
for Josephson networks built on comb graphs.
\end{abstract}

\maketitle

\section{Introduction}

Quite recently it has been evidenced that quantum states may support new
kind of orders, which cannot be characterized by broken symmetries and,
thus, cannot be described by the conventional Ginzburg-Landau
theory \cite{wen02}. Quantum orders may be viewed as
the pertinent description of the pattern of the  quantum entanglement in
a quantum many-body ground-state. Quantum or topological
orders have already been extensively used in the analysis of Fractional
Quantum Hall systems \cite{wen90}, leading to an elegant explanation of
their robust (against a weak but, otherwise, arbitrary perturbation)
ground-state degeneracy and evidencing the intimate connection between
the ground-state and the statistics of quasi-particles. The robustness
of the ground-state degeneracy makes quantum ordered states very
relevant candidates for the engineering of quantum devices
naturally taming  the intrinsic decoherence of other solid state
devices \cite{makhlin01}. Very recent studies hint to the new and
exciting possibility that the topology of Josephson junction networks
(JJN) may be crucial for inducing novel and unexpected macroscopic quantum 
phenomena \cite{burioni00} as well as for opening \cite{ioffe02} to the
possibility for an explicit realization of
a quantum ordered state.

In this paper we shall focus our attention on the recently discovered
possibility that JJN built on suitable graphs may support new and
interesting quantum macroscopic states induced only by a pertinent
engineering of the geometry and topology of the graph supporting the
network \cite{burioni00,burioni01}. In particular, we shall consider JJN
built on comb-like graphs, evidencing the fact that - already for this
very simple graph topology - the array exhibits quantum macroscopic
coherence at low temperatures. After discussing the theoretical
possibility of a spatial Bose-Einstein condensation (BEC) of Cooper
pairs for a JJN built on a comb-graph, we shall propose here a simple
experiment which could enable to observe BEC in these systems.

A network of superconducting classical Josephson junctions is usually described
by the Hamiltonian \cite{fazio01}:
\begin{equation}
H_{JJ}=- J_0 \sum_{\langle x,y; \, x',y' \rangle} \cos(\phi_{x,y} -
\phi_{x',y'}).
\label{qphase}
\end{equation}
On each site of the network there is a superconducting grain and the
junctions are located between neighboring sites with Josephson energy
$J_0$; $\phi_{x,y}$ is the phase of the superconducting order parameter
at site labeled by $(x,y)$. The parameter $J_0$ is connected to the
critical Josephson current $I_c$ by the relation $I_c= 2e J_0 / \hbar$
 \cite{barone82}. The sum  $\langle x,y; \, x',y' \rangle$ runs over all
the distinct nearest-neighbors pairs. The Hamiltonian
(\ref{qphase}) is closely related to the Hamiltonian describing
non-interacting bosons hopping on a generic network
\begin{equation}
H= -t\sum_{\langle x,y;\, x',y'\rangle} \left( \hat{a}^{\dag}_{x,y}
\hat{a}_{x',y'} + H.C.\right) .
\label{h2}
\end{equation}
In Eq. (\ref{h2}), $t$ is the hopping parameter while
$\hat{a}_{x,y}^{\dag}$ ($\hat{a}_{x,y}$) is the creation (annihilation)
operator for bosons;  $\hat{n}_{x,y}= \hat{a}_{x,y}^{\dag}\hat{a}_{x,y}$
is the number operator at site $(x,y)$. The filling, 
i.e., the average number of particles
per site, is defined as $f=N_T/L^2$, where $N_T$ is the total
number of bosons and $L^2$ is the number of sites. When $f \gg 1$ and
the fluctuations of the particle numbers per site are much smaller than
$f$, one can safely substitute the operator $\hat{a}_{x,y}$ with
$\sqrt{N_{x,y}} \: e^{i \phi_{x,y}}$, where $N_{x,y}$ is the number 
of particles at the site $(x,y)$ 
 \cite{vanotterlo93,anglin01}.
As a result, the Hamiltonians (\ref{qphase}) and (\ref{h2}) may be
regarded as equivalent, provided that
\begin{equation} J_0\approx 2 t \sqrt{N_{x,y}\: N_{x',y'}} \approx
2 t f. 
\label{Jos-ij}
\end{equation}

The plan of the paper is the following: in Sec. II we analyze the
spectrum of bosons hopping on a comb-shaped graph. There we
shall also use a variational approach aimed to elucidate
how a mere change in the
network's topology is crucial for explaining the emergence of localized states
in the single-particle spectrum of free bosons hopping on a pertinent graph.
In Sec. III we analyze the thermodynamic properties of
non-interacting bosons hopping on a comb-graph: we compute the critical
temperature $T_c$ and the inhomogeneous spatial distribution of bosons
over the array as a function of the scaled temperature $T/T_c$.
In Sec. IV we propose a method to detect
BEC in a comb-shaped JJN: for this purpose we shall 
determine the Josephson critical current $I_c$ as a function
of the scaled temperature $T/T_c$ and of the location of the junction.

\section{Topology induced localization for bosons hopping on a graph}

In this section, we shall illustrate how a mere change in the topology
of a lattice may affect the spectrum of a quantum mechanical system by
analyzing the properties of non-interacting bosons hopping on a
comb-shaped network.

\begin{figure}
\includegraphics[scale=.32]{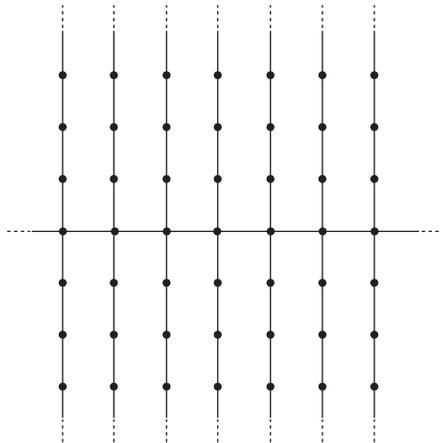}
\caption{ \label{fig1} Comb graph.}
\vspace{8mm}
\end{figure}

A comb graph (see Fig. \ref{fig1}) is composed of one-dimensional chains
({\em fingers}) grafted periodically on a linear chain ({\em backbone}).
Each site of the comb can be naturally labeled introducing two integer
indices $(x, y)$, where $x$ labels the different fingers and $y$
represents the distance from the backbone. Each site on the finger is
linked to two neighbors whereas each site of the backbone has four
neighbors.
\noindent
The Hamiltonian (\ref{h2}) on a generic graph can be written as:
\begin{equation}
H=-t\sum_{x,y;\:x',y'}
A_{x,y;\:x',y'}\: \hat{a}^{\dag}_{x,y} \hat{a}_{x',y'} \; .
\label{pure_hopping}
\end{equation}
The topology of the network is fully described by the adjacency
matrix $A_{x,y;\:x',y'}$ which is equal to $1$ if $(x,y; \: x',y')$  is
a link and $0$  otherwise. The single-particle energy spectrum $\sigma$
on a $L \times L$ comb array may be found by
solving the eigenvalue equation \cite{burioni00,burioni01}:
\begin{equation}
-t \sum_{x',y'} A_{x,y;\:x',y'}\: \psi_E(x',y')= E
\psi_E(x,y) .
\label{eigen_eq}
\end{equation}
The spectrum $\sigma$ is formed by $L^2$ states and 
it is divided in four regions:
$\{E_0\}$, $\sigma_{-}$, $\sigma_{0}$ and $\sigma_{+}$ \cite{burioni00}.
$E_0$ is the ground-state energy
which, with periodic boundary conditions and in the thermodynamic
limit $L \to \infty$, is given  by
$E_{0}=-2 \sqrt{2}t$. $\sigma_0$ is the part of
the spectrum corresponding to delocalized states with energies between
$-2t$ and $2t$; the density of states of $\sigma_{0}$ is
\begin{equation}
\rho(E)= \frac{1}{\pi \:
\sqrt{4t^2-E^2}}\:,
\label{rho_catena}
\end{equation}
just as for a particle hopping in a linear chain. The so-called hidden
spectrum \cite{burioni00} is given by the union of $\sigma_{-}=\{ -2t
\sqrt{1+\cos^2{(k)}}\}$ and $\sigma_{+}=\{ 2 t \sqrt{1+ \cos^2{(k +
\pi/2)}}\}$, where $k=2\pi n/L$ is the wave vector along the backbone
introduced by the Fourier transform of the $x$ coordinate and $n=1,
\dots,(L-1)/4$; the hidden spectrum does not contribute to the
normalized density  of states in the thermodynamic limit since the
number of these states is $L$ \cite{burioni00}. The ground-state and
the eigenstates belonging to $\sigma_{-}$ and $\sigma_{+}$ - due to the
particular topology of the array - are localized along the backbone and
exhibit an exponential decay in the direction of the fingers. Omitting
the normalization factor, the ground-state wavefunction is given 
by \cite{burioni00}
\begin{equation}
\label{ground_state_comb_2} \psi^{comb}_{E_0}(x,y)=e^{- \vert y \vert
/\xi_0} ,
\end{equation}
where $\xi_0$ satisfies the transcendental equation
$\sinh(1/\xi_0)-{\rm{coth}}[L/(2 \xi_0)]=0$; the associated
eigenvalue is $E_0=-t(e^{-1/\xi_0}+e^{1/\xi_0})$. 
As we shall see in the next section, bosons are allowed to spatially
condense in this localized state.
\begin{figure}
\includegraphics[scale=.32]{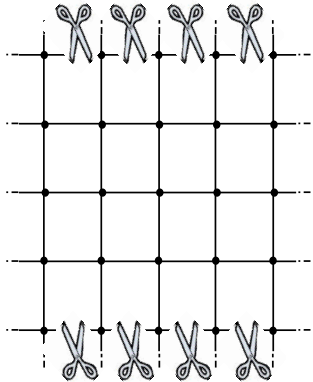}
\caption{ \label{fig2} From square lattice to comb.}
\vspace{8mm}
\end{figure}

It is instructive to provide a more intuitive path to evidence {\it how}
a mere change in the topology of a network may give rise to the
localized ground-state for bosons hopping on a comb graph. For this
purpose one may start from a square lattice and imagine to {\it
systematically} cut the bonds linking the sites of the square lattice in
the $x$-direction at a chosen value of the $y$ coordinate (see Fig.
\ref{fig2}); if one cuts all the bonds in the $x$ direction but the ones
at the origin $y=0$, one obtains the comb lattice. Starting from a
square lattice with $(2M+1) \times (2M+1)$ sites, the procedure to
follow is to remove all the links in the $x$ direction at distance
$\vert d \vert=  M $ from the origin and then \textit{gradually} cut
stripes of links at $\vert d\vert= M-1, \dots, 1 $. As a result,
one gets $M$ different arrays, each described by a different adjacency
matrix $A^{m}_{x,y; \: x',y'}$, where $2m$ is the number of cuts and
$(2M+1)-2m \equiv 2 b+1$ is the number of remaining backbones
($b=M-m$). For the square lattice one has:
\begin{equation}
 A^{sq.}_{x,y;\:x',y'}=  (\delta_{x,x'+1}+\delta_{x,x'-1})\cdot 
\delta_{y,y'}
+\delta_{x,x'} \cdot (\delta_{y,y'+1}+\delta_{y,y'-1}),
\label{square}
\end{equation}
for a square with $2m$ cuts the adjacency matrix is, for $m \ge 1$, given by
\begin{equation}
 A^{m}_{x,y;\:x',y'}= A^{sq.}_{x,y;x',y'} -2  \sum_{q=0}^{m-1}
(\delta_{x,x'+1}+\delta_{x,x'-1})\delta_{y,y'}\delta_{y,M-q}.
\label{cuts}
\end{equation}

To determine the energy of the ground-state as a function
of the number of cuts, it is most convenient to use a variational
approach. For this purpose one may use a trial ground-state wavefunction
$\psi_{\xi}(x,y)$ and then minimize the value of the energy $E(\xi,b;M)$
with respect to the parameter $\xi$. With periodic boundary conditions,
$\psi_{\xi}(x,y)$ should interpolate - as $m$ increases - between the
ground-state wavefunction corresponding to bosons hopping on a square
lattice (a constant with associated eigenvalue $E_0=-4t$) and the
wavefunction (\ref{ground_state_comb_2}), which is the ground-state
wavefunction for bosons hopping on a comb graph.

The simplest choice of the trial ground-state wavefunction taking into
account both limits is given by
\begin{equation}
\label{ansatz_exp}
\psi_{\xi}(x,y)=e^{-\vert y \vert /\xi},
\end{equation}
where $\xi >0 $ is the variational parameter.
The variational energy of the ground-state of bosons hopping on a
lattice with $2m$ stripes of links removed is given by:
\begin{equation}
E(\xi,m;M)=-t \: \frac{\langle{\psi_{\xi}}\vert
A^{m}  \vert \psi_{\xi}\rangle }{\langle{\psi_{\xi}}\vert \psi_{\xi}
\rangle } = -t \: \frac{\sum\limits_{x,y=-M}^{M} \,
\sum\limits_{x',y'=-M}^{M} \psi_{\xi}(x,y) A^{m}_{x,y;x',y'}
\psi_{\xi}(x',y')} {
\sum\limits_{x,y=-M}^{M}\psi_{\xi}(x,y)\psi_{\xi}(x,y)}.
\label{energy}
\end{equation}
\begin{figure}
\includegraphics[scale=.32]{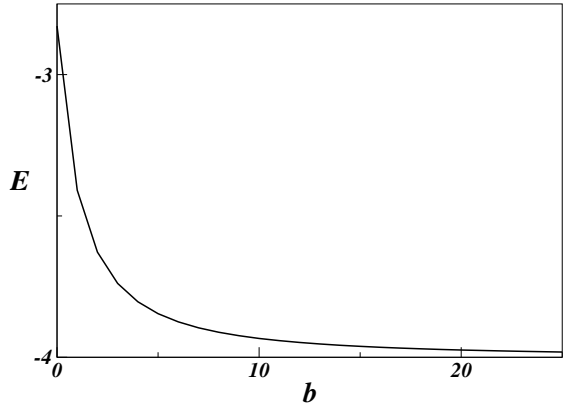}
\caption{ \label{fig3}
Variational energy (\ref{energia_term}), in units of $t$,
of the ground-state of the single-particle
eigenvalue equation (\ref{eigen_eq}) as a function of the number of
backbones $b$ in the thermodynamic limit $M \to \infty$.
When $b=0$, one finds the ground-state
energy of the comb graph $E_0=-2\sqrt{2} t$.}
\vspace{8mm}
\end{figure}
\noindent
As a function of the number of backbones, one has:
\begin{equation}
E(b,\xi;M)= t \:
\frac{e^{-2b/\xi} \left\{ 4e^{(1+2b)/\xi}+e^{ 2M/\xi }\left[ 2-
e^{-2b/\xi} \left( 1+e^{1/\xi} \right)^2 \right]
\right\}}{e^{2M/\xi}+e^{2(M+1)/\xi} -2}.
\label{energia}
\end{equation}
If one minimizes Eq. (\ref{energia}) with respect to
the parameter $\xi$, one obtains the variational ground-state
wavefunction and its associated energy as a function of $b$.

\begin{figure}
\includegraphics[scale=.32]{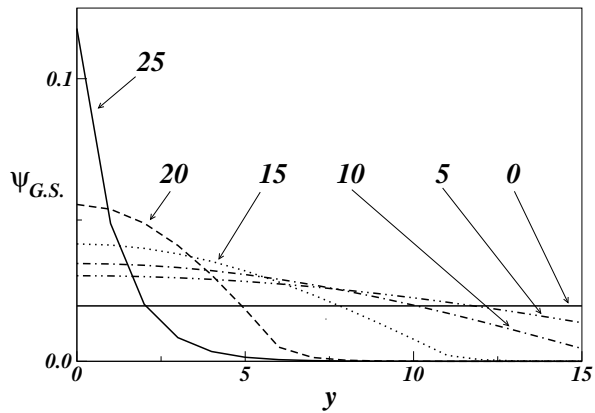}
\caption{ \label{fig4}
The normalized single-particle
ground-state wavefunction $\psi(x,y)$
computed numerically as a function of the distance $y$ from the backbone
for a comb graph with $51 \times 51$ sites and different numbers
of cuts $m=25$, $20$, $15$, $10$, $5$, and $0$. $m=0$ corresponds to the
square lattice and $m=25$ to the comb.}
\vspace{8mm}
\end{figure}

\begin{figure}[t]
\includegraphics[scale=.32]{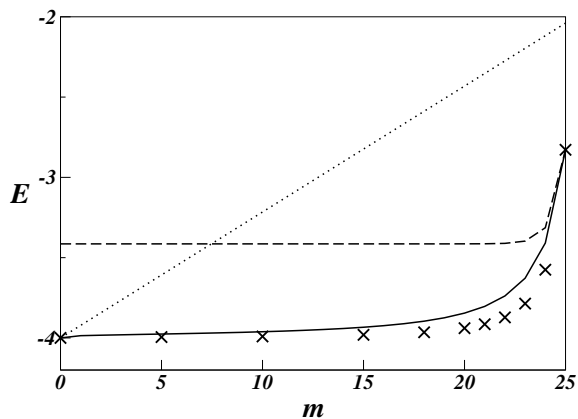}
\caption{ \label{fig5}
Single-particle ground-state energies(in units of $t$)
as a function of the number of
cuts $m$ for a $51 \times 51$ lattice: crosses correspond to the
numerical results, the solid line to the ansatz (\ref{ansatz_exp}).
We also plot the energies of the ground-state of bosons hopping
on a square lattice (dotted line) and on a comb lattice (dashed line).}
\vspace{8mm}
\end{figure}

The results are summarized in Figs. \ref{fig3}-\ref{fig5} and are compared
with the results of a numerical evaluation of the ground-state
wavefunction and its energy. In Fig. \ref{fig3} we plot the variational
energy in the thermodynamic limit which is taken by keeping $b$
(i.e., the number of remaining backbones) constant and performing the
limit $M \to \infty$; the energy is given by
\begin{equation}
E(b,\xi)=  t \, \frac{2 \left[
2 e^{-2b/\xi}- \left( 1+e^{1/\xi} \right)^2
\right]}{1+e^{2/\xi}}.
\label{energia_term}
\end{equation}
As expected, effects due to the network topology clearly emerge if the
number of remaining backbones is a zero-measure set with respect to the
number of stripes in a square lattice.
In Fig. \ref{fig4} we plot the numerical ground-state wavefunction of
bosons hopping on a network with $(51 \times 51)$ sites for different
values of $m$. Figure \ref{fig4} evidences that the simplest trial
wavefunction given in Eq. (\ref{ansatz_exp}) is rather accurate in
describing the shape of the exact wavefunction for a number of
cuts $m \sim M$.
In Fig. \ref{fig5} we compare the energies pertinent to
the ground-states discussed in this section: crosses correspond to the
numerical results and the solid line to the ansatz (\ref{ansatz_exp}). 
We also plot, as a function of the number of cuts $m$, 
the energy of the ground-state 
of bosons hopping on a square lattice: 
for $m=0$ the energy is $-4t$, while for $m = M$ (corresponding to 
the comb lattice) is $-2t$. 
The dashed line is the energy of the ground-state 
(\ref{ground_state_comb_2}) of bosons hopping on a comb lattice as a
function of the number of cuts $m$: for $m=M$
the energy is $-2 \sqrt{2} t$, while for 
$m=0$ the wavefunction (\ref{ground_state_comb_2}) 
is, of course, an excited state. 
Figure \ref{fig5} clearly illustrates how
the localized ground-state of bosons hopping on the comb becomes
energetically more favorable when one increases the number of cuts.

The variational and numerical analysis 
carried out above provides us with an intuitive 
explanation of the appearance of localized states in the spectrum of
a quantum particle hopping on a pertinent graph; furthermore, it clearly
relates this property to the topology change of the network.

\section{Thermodynamics of bosons hopping on a comb graph}

The thermodynamic properties of non-interacting
bosons hopping on a comb graph hint to the possibility of a topology induced
spatial BEC even if $d<2$ \cite{burioni00,burioni01}.
To elucidate this new phenomenon, it is most convenient to introduce
the macrocanonical ensemble to determine the
fugacity $z=e^{\beta(\mu-E_0)}$ as a function of
the temperature \cite{burioni00}: for a lattice with $L \times L$ sites,
the equation determining $z$ is given by
\begin{equation} 
f L^2 \equiv N_T=\sum_{E\in \sigma}
\frac{d(E)}{z^{-1} e^{\beta (E+\sqrt{8}t)}-1}.
\label{filling1}
\end{equation}
In Eq. (\ref{filling1}), $d(E)$ is the degeneracy of each
single-particle eigenstate of the Hamiltonian (\ref{pure_hopping}) and
$\beta=1/k_B T$; the sum in Eq. (\ref{filling1}) is over the entire
spectrum $\sigma$. For free bosons hopping on a comb graph, one
has \cite{burioni00,burioni01,buonsante02}
\begin{equation}
N_T=N_{E_0}(L,T)+N_{\sigma_{-}}(L,T)
+N_{\sigma_{+}}(L,T)+
\int_{E \in \sigma_0} \, dE \,
\frac{L^2 \rho(E)}{z^{-1} e^{\beta (E+\sqrt{8}t)}-1 },
\label{filling}
\end{equation}
where $N_{E_0}(L,T)$, $N_{\sigma_{-}}(L,T)$
and $N_{\sigma_{+}}(L,T)$
denote, respectively, the number of particles at a certain temperature
$T$ in the ground-state and in the two regions $\sigma_{-}$ and
$\sigma_{+}$ of the hidden spectrum. $\rho(E)$, with $E\in \sigma_0$, is
the energy density of states defined in Eq. (\ref{rho_catena}), with
$\sigma_0$ being the region of the spectrum corresponding to delocalized
states. It is also useful to define the number of particles per site in
each part of the spectrum as $n_{E_0}=N_{E_0}/L^2$,
$n_{\sigma_0}=N_{\sigma_0}/L^2$, $n_{\sigma_{-}}=N_{\sigma_{-}}/L^2$ and
$n_{\sigma_{+}}=N_{\sigma_{+}}/L^2$. In the thermodynamic limit, one
has  \cite{buonsante02}
\begin{displaymath}
n_{E_0}(T) = \lim_{L \to \infty} \frac{1}{L^2}
\frac{1}{z^{-1}-1},
\end{displaymath}
\begin{displaymath}
n_{\sigma_{-}}(T)= \lim_{L\to \infty} \frac{2}{L^2}
\sum_{n=1}^{(L-1)/4} \frac{1}{z^{-1} e^{\beta t
[\sqrt{8}-2\sqrt{1+\cos^2(2\pi n/L)}]}-1},
\end{displaymath}
and
\begin{eqnarray*}
n_{\sigma_{+}}(T) &=& \lim_{L\to \infty} \frac{1}{L^2}
\sum_{n=1}^{(L-1)/4} \frac{2}{z^{-1} e^{\beta t
[\sqrt{8}+2\sqrt{1+\cos^2(2 \pi n/L+\pi/2)}]}-1}  \\
&<& \lim_{L\to
\infty}\frac{2}{L}\:
\frac{1}{z^{-1} e^{\beta t( \sqrt{8}+2)}-1 }=0 \quad  \quad \forall \: T.
\end{eqnarray*}
Thus, in the thermodynamic limit, $\sigma_{+}$  is not
macroscopically occupied at any temperature and does not play any role
in describing the thermodynamics of the system.

The last term of the right-hand side of Eq. (\ref{filling})
is the number of bosons in the delocalized (chain-like) states.
The presence of the hidden spectrum changes the behavior of the integral
evaluated in the interval $\{-2t,2t\}$, since it reduces it to the one
describing non-interacting bosons on a linear chain with an impurity in
one of the sites. As a result, letting $z \to 1$, the integral converges
even at finite temperatures making possible a spatial BEC in $d<2$.

\subsection{Critical temperature and condensate fraction}

If one defines $T_c$ as the critical temperature at which BEC occurs,
for any  $T<T_c$ the ground-state is macroscopically filled.
Since, at the critical temperature,
$n_{E_0}(T_c)=n_{\sigma_{\pm}}(T_c)=0 $, the equation allowing to
determine $T_c$ as a function of the parameters $f$ and $t$ reads:
\begin{equation}
\pi f=\int_{-2t}^{2t} \frac{dE}{\sqrt{{4t^2-E^2}} }
\frac{1}{e^{(E+\sqrt{8}t)/(k_{B}T_{c})}-1 }.
\label{critical_T}
\end{equation}

Equation (\ref{critical_T}) can be solved numerically for any
value of $f$.  When $f \gg 1$, one may expand
the exponential in Eq. (\ref{critical_T}) to the first
order in the inverse of the critical temperature $T_c$ getting
\begin{equation}
f \approx \int_{-2t}^{2t}dE
\frac{1}{\pi \: \sqrt{4t^2-E^2}} 
\frac{k_{B}T_{c}}{E+\sqrt{8}t}=\frac{k_B T_c}{2 t}.
\label{critical_t2}
\end{equation}
The critical temperature $T_c$ is then a 
linear function of both $t$ and $f$ given by:
\begin{equation}
T_c \approx \frac{2 t f}{k_B}.
\label{t_c_f}
\end{equation}
Equation (\ref{t_c_f}) has been checked numerically and it is in
excellent agreement with the numerical solution of Eq.
(\ref{critical_T}) for $f \gg 1$, the error being of order $1/f$.
By means of Eq. (\ref{Jos-ij}), one finds that the critical temperature
for the occurrence of BEC is
\begin{equation}
T_c \approx \frac{J_0}{k_B}.
\label{t_c_f_super}
\end{equation}
One may now use Eq. (\ref{t_c_f}) to determine the condensate fraction 
as a function of the scaled temperature $T/T_c$. In the thermodynamic
limit, the number of particles in the delocalized states is given by
\begin{equation}
N_{\sigma_{0}}\Bigg( \frac{T}{T_c} \Bigg)=
\lim_{L \to \infty} L^{2}
\int_{E \in \sigma_0} \rho(E) \frac{dE}{e^{\beta (E+\sqrt{8}t)}-1
}\approx N_T \frac{T}{T_c}.
\label{n_B}
\end{equation}
In the last equation the exponential has been expanded to the
first order in $\beta$: this approximation
holds for $f \gg 1$ and it is in very good agreement with the numerical
evaluation of the integral in Eq. (\ref{n_B}) also in a large
neighborhood below $T_c$. From Eqs. (\ref{filling}) and (\ref{n_B}) one
gets the number of particles in the localized states $N_0=N_{E_0}+
N_{\sigma_{-}}$: the fraction of condensate, for $T<T_c$, is then given
by 
\begin{equation}
\frac{N_0}{N_T} \approx 1 - \frac{T}{T_c}.
\label{n_0}
\end{equation}
For $f$ ranging from $10^3$ to $10^9$, the results provided by 
Eq. (\ref{n_0}) differ from those obtained by the numerical 
evaluation of $N_0$ from Eq. (\ref{filling}) by less than $1 \%$.
Equation (\ref{n_0}) clearly shows that the condensate has dimension
$1$; cigar-shaped one-dimensional atomic Bose condensates support, in
fact, a condensate fraction given by Eq. (\ref{n_0})
 \cite{ketterle96,gorlitz01}.

\subsection{Distribution of bosons along the fingers}

In the following we shall determine the distribution of the bosons
over the comb graph, since - due to the spatial condensation - one
expects an inhomogeneous distribution of the bosons along the
fingers. The average
number of bosons $N_{B}(x,y)$ on a site $(x,y)$ does not depend on $x$ -
due to the translational invariance along the backbone - but only
on the distance $y$ from the backbone, and - at any temperature 
$T$ - is given by: 
\begin{eqnarray}
N_{B}\left(y;T/T_c\right) &=&
N_{E_0}\left(T/T_c\right)\: \vert \psi_{E_0}(y)\vert^2+
\sum_{E_{n}\in \sigma_{-}}N_{\sigma_{-}}\left(E_n;T/T_c\right)
\:\vert \psi_{E_{n}}(y) \vert^2 \nonumber\\ &+& L^{2} 
\int_{E \in \sigma_0} dE \:\rho(E) \: 
\frac{1}{e^{\beta (E+\sqrt{8}t)}-1} \: 
\vert \psi_{E}(y)
\vert^2.
\label{localizzazione}
\end{eqnarray}
In Eq. (\ref{localizzazione}) 
$\psi_{E_0}(y)$ is the wavefunction 
corresponding to the ground-state of the single-particle spectrum and 
$\psi_{E_n}(y)$ are the eigenfunctions corresponding to the energies 
$E_n$ of the hidden spectrum $\sigma_{-}$; $N_{\sigma_{-}}
(E_n)$ is the number of particles with energies $E_n \in \sigma_{-}$
and $N_{E_0}$ is the number of particles in the ground-state.
In the last term of Eq. (\ref{localizzazione}) $\psi_E(y)$ are the 
delocalized eigenfunctions of the eigenvalue equation (\ref{eigen_eq}).

To determine $N_{B}\left(y;T/T_c\right)$, one needs an expression for
$N_{E_0}$ and $N_{\sigma_{-}}$. For this purpose, it is useful to define
the scaled temperature
\begin{equation}
\tau=\frac{T}{T_c}.
\end{equation}
For $\tau \leq 1$ and in the thermodynamic limit, $N_{\sigma_{-}}(E_n,\tau)$
is given by \cite{buonsante02}
\begin{equation}
N_{\sigma_{-}} ( E_n;\tau)= \lim_{L \to \infty} L^2
\frac{2}{\frac{2\sqrt{2}t}{k_B T}(\pi n)^2 +\frac{L^2}{N_{E_0}(\tau)}}
\label{st_nasc_ex}
\end{equation}
and it depends on the number of particles in the ground-state  
$N_{E_0}$.  
Using the fact that $k_B T_c \approx 2 tf$, 
from Eq. (\ref{st_nasc_ex}) it follows that the number of 
particles in the hidden spectrum is given by
\begin{equation}
N_{\sigma_{-}}(\tau)= 
\sum_{n=1}^{\infty} N_{\sigma_{-}} (E_n)=
-N_{E_0}+N_{E_0} 
\sqrt{\frac{\tau}{\sqrt{2}} \frac{N_T}{N_{E_0}}} \;
{\rm{coth}} \, \left[ \sqrt{\frac{\tau}{\sqrt{2}} \frac{N_T}{N_{E_0}}}
\:\right]   . 
\label{st_nasc_ex_tot}
\end{equation}
Requiring that $N_0=N_{\sigma_{-}}+N_{E_0}$ and 
using Eq. (\ref{n_0}), one finds the way to determine 
$N_{E_0}/N_T$ as a function only of the scaled temperature $\tau$:
\begin{equation}
N_T (1 - \tau )=
N_{E_0} 
\sqrt{\frac{\tau}{\sqrt{2}} \frac{N_T}{N_{E_0}}} \;
{\rm{coth}} \,
\left[\sqrt{\frac{\tau}{\sqrt{2}}\frac{N_T}{N_{E_0}}}\:\right].
\label{n_e0}
\end{equation}
Solving Eq. (\ref{n_e0}) and substituting back the value obtained 
for $N_{E_0}$ in Eq. (\ref{st_nasc_ex}) allows for an exact numerical 
evaluation of Eq. (\ref{localizzazione}).

\begin{figure}
\includegraphics[scale=.32]{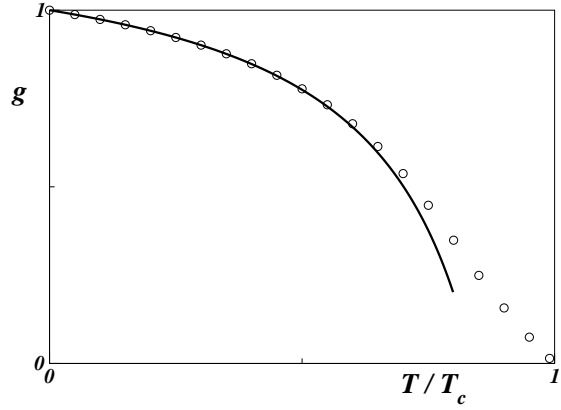}
\caption{ \label{fig6}
The function $g(T/T_c)$ defined in Eq. (\ref{ansatz}): the empty
circles correspond to the numerical solution of the self-consistency
equation (\ref{self_cons1}); the solid line corresponds to the
approximate expression (\ref{self_cons2}).}
\vspace{8mm}
\end{figure}

Conventional wisdom supported by numerical evidence suggests however 
that - apart from a small range of temperatures near $T_c$ - the largest 
contribution to $N_0$ comes from $N_{E_0}$. Thus, it is physically 
appealing to assume the following form for $N_{E_0}$, namely,
\begin{equation}
N_{E_0}=N_T ( 1 - \tau ) g( \tau).
\label{ansatz}
\end{equation}  
In Eq. (\ref{ansatz}), $g(\tau)$ is a function only of the scaled
temperature $\tau$ and parametrizes the contributions to
$N_0$ coming from the states belonging to the hidden spectrum: 
when $g=1$, the condensate is in the ground-state, 
while, for $g=0$, is in the states of the hidden spectrum.
Substituting Eq. (\ref{ansatz}) in Eq.
(\ref{st_nasc_ex}) and requiring 
$N_0=N_{\sigma_{-}}+N_{E_0}$ with $N_0$ given by Eq. (\ref{n_0}), 
leads to a self-consistency equation for $g(\tau)$:
\begin{equation}
g(\tau) \sqrt{\frac{\tau}{\sqrt{2}(1-\tau) g(\tau)}} \;
{\rm{coth}} \, \left[\sqrt{\frac{\tau}{\sqrt{2}(1-\tau) g(\tau)}}
\:\right]=1. \label{self_cons1}
\end{equation}  
For $\tau$ not too close to $1$,
a rather simple approximate solution of 
Eq. (\ref{self_cons1}) is given by
\begin{equation}
g (\tau) \approx 2 - \sqrt{\frac{\tau}{\sqrt{2}(1-\tau)}}
 \; {\rm{coth}} \, \left[\sqrt{\frac{\tau}{\sqrt{2}(1-\tau)}} \:\right].
\label{self_cons2}
\end{equation}  
The error made in using Eq. (\ref{self_cons2}) instead of 
the exact solution of Eq. (\ref{self_cons1}) is within few percents: 
for $\tau \le 0.5$ the error is less than $1 \%$, while for $\tau=0.7$ 
is about $5 \%$. In Fig. \ref{fig6} we plot the function
$g(\tau)$ as obtained from the numerical solution of the
self-consistency Eq. (\ref{self_cons1}) and from the approximate
expression (\ref{self_cons2}).

Upon inserting  Eq. (\ref{self_cons2}) in Eq. (\ref{ansatz}), one has
\begin{equation}
\label{n_0_corr}
\frac{N_{E_0}(\tau)}{N_T} \approx
(1- \tau )  \Bigg
(2 - \sqrt{\frac{\tau}{\sqrt{2}(1-\tau)}}\;
{\rm{coth}} \, \left[\sqrt{\frac{\tau}{\sqrt{2}(1-\tau)}}\:\right]
\Bigg), \end{equation}
and, from Eq. (\ref{st_nasc_ex}), one gets
\begin{equation}
\label{st_nasc_ap}
\frac{N_{\sigma_{-}} ( \tau)}{N_T} = 
\sum_{n=1}^{\infty}
\frac{2}{\frac{\sqrt{2}(\pi n)^2}{\tau}+\frac{1}{(1-\tau) g(\tau)}} \approx
( 1- \tau ) \Bigg\{ 
\sqrt{\frac{\tau}{\sqrt{2} (1-\tau)}} \;
{\rm{coth}} \, \left[ \sqrt{\frac{\tau}
{\sqrt{2} (1-\tau)}}\: \right] - 1 \Bigg\}.
\end{equation}

An explicit analytical form for the number of bosons at site $y$, $N_B(y)$, 
may now be derived. 
The last term in Eq. (\ref{localizzazione}) gives, in fact, 
the contribution coming from the delocalized states: 
for a large network ($L \gg 1$), and far away from the backbone, this number 
is independent from the site index $y$ and equals 
a constant $(N_T/L^2) \tau$. 
Then, using Eqs. (\ref{n_0_corr}) and (\ref{st_nasc_ap}), 
for $\tau<1$, one has
\begin{eqnarray}
N_{B}(y;\tau)&\approx& \lim_{L \to \infty} N_T
\left\{ \frac{1}{L}
\left[
\left( 1- \tau -
\sum_{n=1}^{(L-1)/4} \frac{2}{\frac{\sqrt{2}(\pi n)^2}{\tau}+
\frac{1}{\left(1-\tau\right)g(\tau)}}\right) \frac{e^{-2 \,
{\rm{arcsinh}}(1)\vert y \vert}} {\sqrt{2}} \right.\right. \nonumber
\\
&+& \left.\left. \sum_{n=1}^{(L-1)/4} \frac{2 \cos(2\pi
n/L)}{\frac{\sqrt{2}(\pi n)^2} {\tau}+\frac{1}{(1-\tau)g(\tau)}} \:
\:\frac{e^{-2 \, {\rm{arcsinh}} [\cos(2 \pi n/L)] \vert y
\vert}}{\sqrt{1+\cos^2(2\pi n/{ L})}} \right]+\frac{\tau}{L^2}
\right\} .
\label{rapporto1}
\end{eqnarray}
The exponential behavior typical of the eigenfunctions corresponding 
to the localized states leads, for $\tau < 1$,
to an increase of $N_{B}(y; \tau)$ on the sites near
the backbone while, when $y \gg 1$, the behavior is dominated by the last
term in the right-hand side of Eq. (\ref{rapporto1}). Thus, away from
the backbone, once the filling is fixed, $N_B$ depends only on the
scaled temperature $\tau$ and it is given by
\begin{equation}
\label{rapporto_sempl}
\frac{N_B (y;T/T_c)}{f} \approx \tau \equiv \frac{T}{T_c}.
\end{equation}
In the next section we shall show how this property may be useful to
compute the observable effects induced by the existence of a spatial
BEC in condensed matter systems.

\section{Further outlook: Signature of BEC for Josephson 
networks on a comb}

We proposed so far a theoretical framework establishing the role  played by 
the network's topology in understanding the quantum and thermodynamic
inhomogeneities arising in a system of bosons hopping on a comb graph; the
approach enables one to obtain a simple and explicit analytical
expression for the inhomogeneous distribution of bosons along the comb's
fingers signaling the dramatic effect of the topology induced spatial
BEC evidenced in Refs. \cite{burioni00,burioni01}. With little
modifications our analysis could be carried out also for diverse
network's topologies supporting BEC \cite{burioni01}.

We shall now evidence how to detect the signature of BEC for
a JJN built on a comb graph. For this purpose, one may think
to perform a measurement of the $I$-$V$ characteristic of a
single finger of the JJN and of the critical Josephson current 
along the finger. If one feeds, in fact, an external current 
$I_{ext}$ at the extremities of the finger, one should expect to
observe no voltage unless $I_{ext}$ is larger than the smallest of the
critical currents of the junctions along the finger.
Since, below $T_c$, the critical Josephson current of the finger 
is given by the smallest of the critical currents 
of the junctions positioned 
along the fingers, the measurement of the $I$-$V$ characteristic of 
the finger should provide a measurement of the critical current of the 
junctions at the top of the fingers (i.e., at $y \gg 1$).

To make a definite prediction, one needs to estimate the value of the
Josephson critical current of a single junction of the network as a
function of both the temperature and the distance from the backbone.
Above $T_{c}$ the Cooper pairs, for a large array ($L \gg 1$), 
are uniformly distributed over the network: 
the Josephson critical current is
the same at each junction
\begin{equation} I_c^{A}(y) \approx
t \:(4e/\hbar) \cdot \sqrt{N_{B}(y+1)N_{B}(y)} \approx
J_0 \cdot (2e/\hbar) 
\label{critical_above_Tc}
\end{equation}
where $I_c^{A}$ is the Josephson critical current above 
$T_c$. 
According to Eq. (\ref{Jos-ij}), the Josephson critical current depends
only on the position and on the population of the sites.

\begin{figure}
\includegraphics[scale=.32]{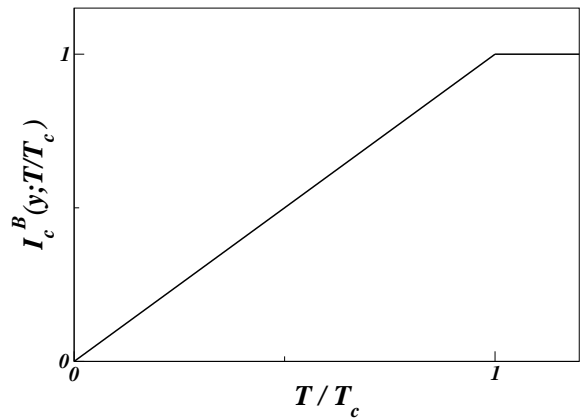}
\caption{ \label{fig7}
Josephson critical current as a function of $T/T_c$ computed at
distances from the backbone $y \gg 1$. $I^{B}_c(y;T/T_c)$ is in units of
$I^{A}_c(y)$ (the critical current of the junctions above $T_c$) and is
therefore equal to 1 for $T\geq T_c$.}
\vspace{8mm}
\end{figure}

The relation between the Josephson critical current of the junction
between the sites $(x,y)$ and $(x,y+1)$ above and below the
critical temperature $T_c$ does not depend on $x$ and it is given by
\begin{equation}
\frac{I^{B}_c(y;\tau)}{I^{A}_c(y)}
\approx \frac{\sqrt{N_{B}(y+1,\tau)N_{B}(y,\tau)}}{f},
\label{rapporto2}
\end{equation}
where $I^{B}_c(y;\tau)$ is the Josephson critical current below $T_c$ at
distance $y$ from the backbone. 
$N_{B}(y,\tau)$ is given by Eq. (\ref{rapporto1}). The result
(\ref{rapporto2}) - for a given $f$ - is independent from the total
number of bosons in the system. Far from the backbone ($y \gg 1$), one
simply has
\begin{equation}
\frac{I^{B}_c(y;\tau)}{I^{A}_c(y)}
\approx \frac{N_{B}(y,\tau)}{f},
\label{rapporto_lontano}
\end{equation}
and, using Eq. (\ref{rapporto_sempl}), one gets
\begin{equation}
\frac{I^{B}_c(y;\tau)}{I^{A}_c(y)}
\approx \tau \equiv \frac{T}{T_c}.
\label{rapporto_lontano_tc}
\end{equation}
$I^{B}_c(y;\tau)$ at the top of the finger can be experimentally 
measured as the critical Josephson current of the finger. 
In Fig \ref{fig7} we plot Eq. (\ref{rapporto_lontano_tc}).

BEC in a JJN built on a comb graph predicts then a rather sharp decrease
of the Josephson critical current for a junction located away from the
backbone and this behavior affects the measurement of the 
$I$-$V$ characteristic along a given finger of the JJN. The linear
dependence exhibited by the solid line in Fig. \ref{fig7} is consistent
with the observation that, in this system, the condensate has dimension
1. We stress that this phenomenon should be already observed
for a classical JJN described by the Hamiltonian (\ref{qphase}).

The slope of the linear plot in Fig. \ref{fig7} provides a direct
estimate for the fraction of condensate $N_{0}/N_T\approx 1-k_B T/J_0$
for a JJN built on a comb graph. In fact, far away from the
backbone ($y \gg 1$), from Eqs. (\ref{n_0}), 
(\ref{rapporto_lontano}) and (\ref{rapporto_lontano_tc}), 
one has that the fraction of condensate is given by 
$1-\big[I^{B}_c(y;\tau)/I^{A}_c(y)\big]$.

To conclude this section, we notice that a similar analysis
applies to a bosonic gas in comb-shaped deep optical
lattice, provided that one defines the Josephson current 
of a single bosonic Josephson junction between two neighbor sites at 
$y$ and $y+1$ as $I_c \approx 2 t \sqrt{N_B(y) N_B(y+1)}$ 
(where $N_B(y)$ is the number of bosons in $y$). Above $T_c$, all the 
$N_B(y)$ will be equal to the filling $f$ at the equilibrium and 
$I_c^{A}=2 tf$. Below $T_c$, the equilibrium values for $N_B(y)$ change
along the fingers according to Eq. (\ref{localizzazione}) and the ratio
between $I_c^{B}(y,\tau)$ and $I_c^{A}$ gives, for $y \gg 1$,
directly the fraction of $noncondensate$ atoms.

\section*{Acknowledgements}

We thank Mario Rasetti, Matteo Cirillo, Carlo Cosmelli, 
and Alessandro Vezzani for the benefit of many stimulating discussions. 
We acknowledge
financial support by M.I.U.R. through grant No. 2001028294.
One of us (P.S.) would like to thank Francesco Guerra for
his long-lasting friendship and continuous 
guidance since the years he was His student at the
University of Naples.

\end{document}